\documentclass[12pt]{iopart}

\usepackage{iopams}
\usepackage{setspace}

\usepackage{amsfonts}   
\usepackage{amssymb}
\usepackage{graphicx}   
\newcommand{\sinc}{sinc}

\begin{document}

\title{Fermi's golden rule: its derivation and breakdown by an ideal model}
\author{Zhang J M$^{1, 2}$ and Liu Y$^{3,4}$}
\address{$^1$ Fujian Provincial Key Laboratory of Quantum Manipulation and New Energy Materials,
College of Physics and Energy, Fujian Normal University, Fuzhou 350007, China}
\address{$^2$ Fujian Provincial Collaborative Innovation Center for Optoelectronic Semiconductors and Efficient Devices, Xiamen 361005, China}
\address{$^3$ LCP, Institute of Applied Physics and Computational Mathematics, Beijing 100088, China}
\address{$^4$ Software Center for High Performance Numerical Simulation, China Academy of Engineering Physics, Beijing 100088, China}

\begin{abstract}
Fermi's golden rule is of great importance in quantum dynamics. However, in many textbooks on quantum mechanics, its contents and limitations are obscured by the approximations and arguments in the derivation, which are inevitable because of the generic setting considered. Here we propose to introduce it by an ideal model, in which the quasi-continuum band consists of equaldistant levels extending from $-\infty $ to $+\infty $, and each of them couples to the discrete level with the same strength. For this model, the transition probability in the first order perturbation approximation can be calculated analytically by invoking the Poisson summation formula. It turns out to be a \emph{piecewise linear} function of time, demonstrating on one hand the key features of Fermi's golden rule, and on the other hand that the rule breaks down beyond the \emph{Heisenberg time}, even when the first order perturbation approximation itself is still valid.
\end{abstract}

%


\section{Introduction}

As an example of Stigler's law of eponymy, the so-called Fermi's golden rule was actually first derived by Dirac \cite{dirac} instead of Fermi, although the title ``golden'' was given by the latter \cite{fermi}. It is no wonder that Fermi considered the rule a golden one in view of its instrumental role in his theory of Beta decay \cite{beta,beta2}. Later development proves that the rule lives up to its name, as it is routinely used in calculating the transition rates or cross sections of various processes \cite{devanathan}. Noteworthily, the usual tunneling rate can also be calculated using the rule \cite{reittu, alex}, if the tunneling process is understood as a transition process under an appropriate perturbation.

As a nontrivial result of the lowest order time-dependent perturbation theory, Fermi's golden rule is introduced in almost every textbook on quantum mechanics. It is about the transition dynamics for such a scenario. Initially, the system is in some eigenstate $|b\rangle $ of an unperturbed Hamiltonian $H_0 $ with eigenenergy $E_b$. Besides the level $|b\rangle $, $H_0 $ has a quasi-continuum $\{ |n\rangle  \}$ with eigenenergies $\{ E_n \}$. It is assumed that $E_n$ is non-degenerate and increases with $n$, and that the quasi-continuum covers an interval $[a,  b]$, to which $E_b$ belongs, i.e., $a< E_b < b$. Then at $t= 0 $ a perturbation $V $ is turned on, which couples $|b\rangle $ and $ \{ |n\rangle  \} $ (and possibly, states within the continuum band too, say $|n_1\rangle $ and $|n_2 \rangle $) with strength $g_n = \langle n |V | b \rangle $. It is assumed that $g_n$ is a slowly varying function of $n $, and therefore, it is legitimate to introduce a continuous function $g(E)$ such that $g_n= g(E_n)$. Because of the newly introduced coupling, the system transits towards the quasi-continuum. Fermi's golden rule then states that, in the first order perturbation theory, the probability $P $ of finding the system in the continuum  grows linearly in time, and the rate of increase, the so-called transition rate, is of the expression
\begin{eqnarray}\label{fgr}
  w = \frac{d P}{d t} = \frac{2\pi}{\hbar } |g (E_b )|^2 \rho(E_b) .
\end{eqnarray}
Here $\rho(\cdot )$ is the density of states of the quasi-continuum. It is defined as $\rho(E) dE $ is the number of levels in the interval $[E, E+ d E]$.

In many textbooks on quantum mechanics, formula (\ref{fgr}) is derived for a generic case. In other words, no concrete constraint is placed on the quasi-continuum spectrum $\{ E_n  \}$ or the couplings $\{g_n  \}$, except for the conditions above, which are either implicitly or explicitly assumed. The generality of this approach is definitely appreciable, but exactly because of its generality, many approximations and arguments (within the first order perturbation framework) are inevitable, which obscure the contents and limitations of the rule, especially for a novice. For example, in deriving (\ref{fgr}), in most if not all textbooks \cite{diracbook, messiah, landau,cohen, weinberg, sakurai, zeng, merzbacher, ballentine, baym}, the limit of $ t \rightarrow \infty $ is taken. However, intuitively, one would not expect the first order perturbation theory to hold any more in the long time limit of $t \rightarrow \infty $. Actually, to be consistent with the first order perturbation theory, the condition $P  \ll 1 $ should be satisfied, and the linear behavior of (\ref{fgr}) cannot last forever. Although this dilemma is just superficial (the $t\rightarrow \infty $ limit is essentially achieved already for a short and finite $t$), it does baffle the beginners or even experts. At this point, we would like to echo Stedman \cite{stedman}, ``\emph{It is the author's experience that the majority of physicists who use the golden rule have never justified it to their own satisfaction}''.

It is therefore desirable to have a model for which the calculation can be done as rigorously and as straightforwardly as possible. Such a model indeed exists, and as a matter of fact, it has been known for decades \cite{stey}. Yet, unfortunately, to the best knowledge of the author, it has not been introduced into any quantum mechanics textbook, neither in the main text nor in the exercise part. The model is very simple---in hindsight, it can actually be motivated by the expression (\ref{fgr}). The transition rate  $w$ is proportional to the local values of the density of states and the coupling strength squared. Hence, in the model, the level spacing $E_{n+1}- E_n$ and the coupling $g_n $ are simply taken as constant. For this model, the transition probability in the first order approximation can be calculated rigorously for arbitrary time $t$. It is actually a piecewise linear function of time, showing kinks with a period of $t_{H}$, the Heisenberg time associated with the spectrum. Therefore, on one hand, the approximations and arguments in the general case, in particular, the $t\rightarrow \infty $ limit, are avoided; on the other hand, it presents an example demonstrating vividly that Fermi's golden rule breaks down beyond the Heisenberg time, even when the first order approximation itself is still valid. As far as the author knows, the Heisenberg time as an upper bound for the validity domain of Fermi's golden rule is pointed out only by Baym \cite{baym}. In many books, it is simply given by the non-depletion condition $t \ll 1/w $ \cite{diracbook, messiah, cohen, merzbacher}.

In the following, we shall first present the general formalism of the first order time-dependent perturbation theory and derive Fermi's golden rule in the generic case in Sec.~\ref{genericsection}. Then, we proceed to consider the special case of the ideal model in Sec.~\ref{idealsection}. It is interesting that the model is not only solvable up to the first order approximation, but can be solved exactly by summing up all orders of terms. For completeness, this is done in Sec.~\ref{fullsection}.

\section{The generic case}\label{genericsection}
Let us first review the derivation of (\ref{fgr}) in the generic case. The time dependent Schr\"odinger equation is
\begin{equation}\label{seq}
  i \hbar \frac{\partial |\psi \rangle }{\partial t } = (H_0 + \lambda V) |\psi \rangle .
\end{equation}
The initial condition is $|\psi(t=0 ) \rangle=|b\rangle $. Here we have introduced a control parameter $\lambda$, whose value will be set to $1$ in the end. Expand the wave function $| \psi(t;\lambda ) \rangle$ as a power series of $\lambda$,
\begin{equation}\label{expan1}
 | \psi(t;\lambda ) \rangle = \sum_{s=0}^\infty \lambda^s |\psi_{s} (t) \rangle.
\end{equation}
By the initial condition of $| \psi \rangle$, we have $| \psi_{0} (0) \rangle =|b\rangle $ and $ | \psi_{s}(0) \rangle= 0$ for $s\geq 1 $. Plugging (\ref{expan1}) into (\ref{seq}), and equating the coefficients of the powers of $\lambda$, we get up to $s=1$,
\begin{eqnarray}
  i\hbar \frac{\partial | \psi_{0} \rangle }{\partial t } &=& H_0 | \psi_0 \rangle ,   \label{zeroth}  \\
  i \hbar \frac{\partial |\psi_1 \rangle}{\partial t } &=& H_0 |\psi_1 \rangle+ V |\psi_0 \rangle .  \label{1st}
\end{eqnarray}
From (\ref{zeroth}) we solve $| \psi_0(t) \rangle= e^{-i E_b t/\hbar } |b\rangle $. Substituting this result into (\ref{1st}), and projecting both sides of (\ref{1st}) onto $|n\rangle $,  we get
\begin{equation}
  i\hbar \frac{\partial }{\partial t } \langle n | \psi_1 \rangle = E_n \langle n | \psi_1 \rangle + g_n e^{-i E_b t /\hbar} .
\end{equation}
By Duhamel's principle \cite{duhamel}, we then easily solve
\begin{equation}
 \fl \quad \quad \quad \quad  \langle n | \psi_1(t) \rangle = \frac{1}{i \hbar} \int_0^t d \tau  g_n e^{-i E_b \tau /\hbar } e^{-i E_n (t- \tau)/\hbar } = g_n \frac{ 1 - e^{i (E_n - E_b )t/\hbar } }{E_n - E_b } e^{-i E_n t /\hbar } .
\end{equation}
Therefore, to the first order approximation, the probability of finding the system in the quasi-continuum band is
\begin{eqnarray}\label{1stappro}
  P =\sum_n | \langle n | \psi_1(t) \rangle |^2  &=& \sum_n |g_n|^2 \frac{4 \sin^2 ((E_n- E_b )t / 2 \hbar)}{(E_n - E_b )^2} \nonumber \\
  &=&   \frac{2t}{\hbar } \sum_n |g_n|^2 \frac{ \sin^2 ((E_n- E_b )t / 2 \hbar)}{(E_n - E_b )^2 (t/2\hbar )} .
\end{eqnarray}
Under certain condition, the summation can be approximated by an integral
\begin{equation}\label{integral}
  P = \frac{2t}{\hbar }\int_a^b d \epsilon \rho(\epsilon) |g(\epsilon )|^2  \frac{ \sin^2 ((\epsilon- E_b )t / 2 \hbar)}{(\epsilon  - E_b )^2 (t / 2 \hbar )}.
\end{equation}
The weight function (as a function of $\epsilon $) in the integral
\begin{equation}\label{weight}
 \frac{\sin^2 ((\epsilon- E_b )t / 2 \hbar)}{(\epsilon  - E_b )^2(t / 2 \hbar)}
\end{equation}
consists of lumps whose width scales as $1/t$, and whose height scales as $t$. In the limit of $t\rightarrow \infty$, it converges to the delta function $ \pi \delta (\epsilon - E_b )$. In this limit,
\begin{equation}\label{infinity}
  P = \frac{2\pi t}{\hbar } \rho(E_b) |g(E_b )|^2  .
\end{equation}
We then get the golden rule (\ref{fgr}). Here we have invoked two approximations ensuing (\ref{1stappro}), which is an approximation by itself. From (\ref{1stappro}) to (\ref{integral}), in replacing the summation by an integral, the sampling step-length (i.e., level spacing $E_{n+1} -E_n $) should be much smaller than the characteristic length of variation of the function, which is $2\pi \hbar / t$. That is,
\begin{equation}\label{upper}
  t \ll 2\pi \hbar \rho(E_b) .
\end{equation}
From (\ref{integral}) to (\ref{infinity}), in replacing the weight function (\ref{weight}) by the delta function $\pi \delta (\epsilon - E_b)$, the width of the lumps of the former should be much smaller than the width of the interval $[a, b]$, or more accurately,
\begin{equation}\label{lower}
   \frac{2\pi \hbar }{ \min \{ |a- E_b |, |b- E_b | \}} \ll t .
\end{equation}
Both conditions of (\ref{upper}) and (\ref{lower}) are within the first order perturbation approximation itself, and can be satisfied simultaneously. We note that while condition (\ref{lower}) is mentioned in many books \cite{diracbook, messiah, cohen, baym}, condition (\ref{upper}) is pointed out only by Baym \cite{baym}. As we shall see in the ideal model below, the right hand side of (\ref{upper}) defines sharply the boundary beyond which Fermi's golden rule breaks down.

\section{The ideal model}\label{idealsection}
\begin{figure}[tb]
\includegraphics[width= 0.4\textwidth ]{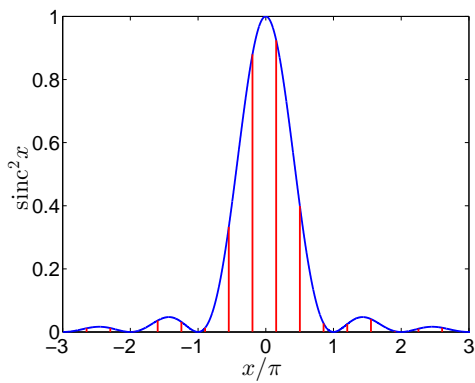}
\caption{(Color online) Periodic sampling  and summation of the function $\sinc^2 x \equiv \sin^2 x / x^2$, which is the essential part in the definition of $W_\alpha (T )$ [see (\ref{w2})]. The sampling period is $T$, and the shift is determined by $\alpha $. In this figure, $\alpha = 3/7$.}
\label{sampling}
\end{figure}

In retrospect, the ideal model can be motivated by the expression of the golden rule (\ref{fgr}) or (\ref{infinity}). The transition rate is proportional to the local values of the density of state $\rho(E)$ and the square of the coupling $|g(E)|^2$ at $E=E_b$. In the ideal model, both functions are simply taken as constant. That is, $E_n  = n \Delta $ with $\Delta $ being the spacing between two adjacent levels, and $g_n = g$.
More specifically, the unperturbed Hamiltonian and the coupling are, respectively \cite{stey},
\numparts
\begin{eqnarray}\label{model}
H_0 &=& E_b |b \rangle \langle b | + \sum_{n=-\infty}^{\infty} n \Delta | n\rangle \langle n | , \label{modelh0} \\
V  &=& g\sum_{n=-\infty}^{\infty} ( |b \rangle \langle n | +  |n \rangle \langle b | ). \label{modelv}
\end{eqnarray}
\endnumparts
Note that the quasi-continuum extends from $-\infty $ to $+\infty$, and we have assumed that the coupling strength $g$ is real, which is always achievable by adjusting the phases of the levels $| n \rangle $.
This model can be approximately realized with a two-level atom in a \emph{multi-mode} cavity \cite{parker,meystre,ligare95, ligare02}. For detailed derivation see \cite{ligare95, ligare02}. The basic idea is that, under the rotating wave approximation, we have a multi-mode Jaynes-Cummings model, in which the total excitation number (the excitation number of the atom is 1 if it is in the excited level and 0 if it is in the ground state, and the excitation of each cavity mode is the Fock number) is conserved. In the single excitation subspace, we have the level-continuum structure, in which the discrete level $|b\rangle $ corresponds to the atom in the excited state and all cavity modes empty, while each $|n\rangle $ in the continuum corresponds to the atom in the ground state and the $n$th cavity mode being excited. The mode spacing and the atom-mode couplings (if the atom is appropriately located) are, to a good extent, constant. Only those modes near resonant with the atomic resonance frequency participate significantly in the dynamics. Therefore, it is legitimate to assume that there are infinite number of cavity modes extending from $-\infty$ to $+\infty $.

For the concrete model of (\ref{modelh0}) and (\ref{modelv}), the general formula (\ref{1stappro}) reduces to periodic sampling and summation of the function $\sinc^2 x \equiv \sin^2 x / x^2 $. Specifically, introducing the dimensionless time $T \equiv  \Delta t /2 \hbar $ and the offset parameter (here $\lfloor \cdot \rfloor$ is the floor function)
 \begin{eqnarray}\label{defalpha}
 \alpha = E_b/\Delta - \lfloor {E_b /\Delta } \rfloor,
 \end{eqnarray}
which characterizes the location of $E_b$ relative to the quasi-continuum spectrum, the transition probability can be written as
\begin{equation}\label{w1}
  P = \left( \frac{4g^2}{\Delta^2 } \right )  W_\alpha (T) .
\end{equation}
Here the function $W_\alpha $ is defined as
\begin{eqnarray}\label{w2}
W_\alpha (T) \equiv  T^2 \sum_{m =-\infty}^{\infty } \sinc^2 [(m-\alpha )T],
\end{eqnarray}
where the infinite summation is a very regular one---it samples the $\sinc ^2 x $ function uniformly with the period given by $T$ and the offset determined by $\alpha  $ (see figure \ref{sampling}). It is apparent that $W_\alpha = W_{-\alpha} = W_{1-\alpha}$, i.e., $W_\alpha$ is an even and periodic function of $\alpha$. Note that in (\ref{w1}), the time dependence is only through the function $W_\alpha$ and the coupling strength $g$ appears only in the prefactor.

The primary concern is to calculate $W_\alpha $. For this purpose, we have the standard tool of Poisson summation formula \cite{Mussardo}. By this formula, a periodic sampling and summation of a function $f(x)$,
\begin{eqnarray}\label{realsum}
  I = \sum_{n=-\infty}^{+\infty } f(a+ n T)
\end{eqnarray}
can be converted to a weighted periodic sampling of its Fourier transform
\begin{eqnarray}
F(q)= \int_{-\infty}^{+\infty } dx f(x) e^{-iq x}.
\end{eqnarray}
That is,
\begin{equation}\label{momsum}
I = \frac{1}{T} \sum_{n=-\infty}^{+\infty } F\left(\frac{2\pi n }{T}\right) \exp\left( \frac{i 2\pi n a }{T} \right).
\end{equation}
In our case, we need to calculate the Fourier transform of the function $\sinc^2 x$. For those who are familiar with the Fraunhofer diffraction of a single slit \cite{hecht}, it is a basic mathematical fact that the Fourier transform $F_1$ of the function $\sinc ( x) $ is the window function \cite{around, window}. Namely,
\begin{eqnarray}
F_1 (q ) =  \cases{ \pi  & if $ |q|\leq 1, $ \\ 0 &  if $ |q| > 1 . $ \\  }
\end{eqnarray}
By the convolution theorem, the Fourier transform $ F_2$ of $\sinc^2 x$ is then simply the self-convolution of $F_1$,
\begin{eqnarray}
F_2(q ) = \frac{1}{2\pi } \int_{-\infty}^{+\infty} d p F_1(p) F_1(q-p)= \cases{  \frac{\pi}{2}  (2-|q|) , & $ |q| \leq 2 $ , \\ 0, & $ |q| > 2  , $  \\ }
\end{eqnarray}
which is a triangle function nonvanishing only on the support $[-2, 2]$. The fact that the Fourier transform of $\sinc^2 x$ has only a finite support is actually a consequence of the Paley-Wiener theorem \cite{rudin}. The function $\sinc^2 x$ is an entire function and is of exponential type 2 \cite{exponentialtype}. Hence, by the Paley-Wiener theorem, its Fourier transform is supported on $[-2, 2]$.

The fact that $F_2$ is nonvanishing only on a finite interval means that invoking the Poisson summation formula really simplifies the original summation problem in (\ref{w2}), because in (\ref{momsum}) there will be only a \emph{finite} number of nonzero terms. For example, if $0< T \leq  \pi $, only the $ n=0 $ term is nonzero, and consequently, in this interval,
\begin{eqnarray}\label{linear}
  W_\alpha (T) &=& \pi T ,
\end{eqnarray}
regardless of the value of $\alpha $. In terms of the real time $t$, the result is that, for $0< t<t_{H} \equiv  2\pi\hbar /\Delta $, the Heisenberg time,
\begin{eqnarray}
  P &=& \frac{2\pi}{\hbar } \left( \frac{g^2}{\Delta } \right ) t .
\end{eqnarray}
This is nothing but Fermi's golden rule. We have obtained it without using any approximation or argument. The fact that it is independent of the parameter $\alpha $ is in accord with the coarse-graining spirit implicitly assumed in the usual derivation of the rule. Here we have introduced the notion of the Heisenberg time. The name apparently comes from the usual time-energy uncertainty principle---to resolve two levels spaced by $\Delta$, we need to wait for a period on the scale of $1/\Delta$.

However, in the more general case of $m \pi < T \leq (m+1) \pi $, there are $2m +1$ nonzero terms in the summation. It is straightforward to get ($\theta \equiv  2\pi \alpha $)
\begin{eqnarray}\label{wfull}
W_\alpha (T)
&=&  \pi \left( \sum_{n=-m}^{m} \exp (i n \theta ) \right)  T  - 2 \pi^2 \sum_{n=1}^m  n \cos ( n \theta) \nonumber \\
&=& \pi  \frac{\sin\frac{(2m+1)\theta}{2}}{\sin \frac{\theta}{2}  }  T - 2\pi^2 \frac{\partial}{\partial \theta} \left( \sum_{n=1}^m \sin n \theta \right)   \nonumber \\
&=& \pi  \frac{\sin\frac{(2m+1)\theta}{2}}{\sin \frac{\theta}{2}  } (T-m \pi ) + \pi^2  \frac{\sin^2  \frac{m \theta}{2}}{\sin^2 \frac{\theta}{2}} .
\end{eqnarray}
This formula was first obtained by Kyr\"ol\"a using a different method \cite{kyrola}.
We see that $W_\alpha$ is still a linear function of $T $ on the interval $[m \pi, (m+1)\pi ]$, but its slope now depends on both the interval and the offset parameter $\alpha$. Therefore, $W_\alpha $ is a piecewise linear function of $T$ and has kinks at $T= m \pi $ (correspondingly, $t = m t_{H}$) periodically. The point is that, it is continuous but nonsmooth. Apparently, by making $g$ sufficiently weak, one can make $P$ arbitrarily small at an arbitrary time $t$. Therefore, for a sufficiently small $g$, the first order perturbation approximation can be valid far beyond the critical time $t_{H} $. Yet, Fermi's golden rule is no longer valid there.

\begin{figure}[tb]
\includegraphics[width= 0.4\textwidth ]{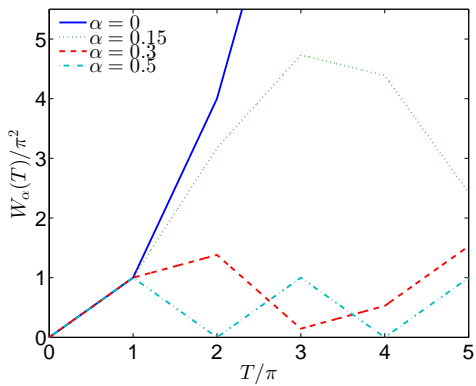}
\caption{(Color online) The function $W_\alpha (T )$ [see (\ref{w2})] for four different values of $\alpha $. Each curve is piecewise linear and all the curves coincide in the first interval of $0\leq T \leq \pi $. Note the two particular cases of $\alpha = 0 $ and $\alpha = 0. 5$. For the former, $W_\alpha$ diverges quadratically with $T$; for the latter, $W_\alpha$ returns to zero (as a phenomenon of collapse and revival) repeatedly. }
\label{Wafig}
\end{figure}

In figure \ref{Wafig}, the function $W_\alpha (T)$ is plotted for four different values of $\alpha $. Before $t_{H}$, all curves coincide as (\ref{linear}) says; but beyond $t_{H}$, the curves fan out and have completely different trajectories. Except for the case of $\alpha = 0$, namely when $E_b$ is degenerate with a certain level $|n\rangle $, $W_\alpha$ shows collapses and revivals. Actually, by (\ref{wfull}), at $T= m \pi$,
\begin{eqnarray}
  W_\alpha (m \pi )  = \pi^2  \frac{1- \cos m \theta }{1- \cos \theta }.
\end{eqnarray}
Thus for a generic value of $\alpha$, $W_\alpha (m \pi )$ is an almost periodic function of $m$.

We note that the piecewise linear behavior of the transition probability as illustrated in figure \ref{Wafig} has been numerically observed in the contexts of photoexcitation of a molecule \cite{kyrola}, spontaneous decay of a two-level atom in a multi-mode optical cavity \cite{parker,meystre}, and transition dynamics of a Bloch state in a periodically driven tight binding model \cite{scienceopen}. In the third of these systems, the model (\ref{modelh0}) and (\ref{modelv}) is actually not realized exactly. Except for the superficial time dependence, the perturbation $V$ contains intra-continuum terms, and the level spacing is not strictly constant. But to the first order of $g$, the intra-continuum couplings do not enter the dynamics, and only the Bloch states near resonance play an important in the dynamics, for which the equaldistant level spacing condition is a good approximation. Therefore, the perturbation theory above still holds.


\section{Exact solution of the transition problem}\label{fullsection}

The transition problem for the ideal model can actually be solved exactly. This was originally done by Stey and Gibberd using the Laplace transform technique \cite{stey} and later by Lefebvre and Savolainen using the memory function method \cite{lefebvre}. In the following, a solution in line with the perturbation expansion (\ref{expan1}) will be presented.

To determine $|\psi_s(t)\rangle $ in (\ref{expan1}), let us transform into the interaction picture \cite{sakurai}. Define $|\psi (t) \rangle= e^{-i H_0 t /\hbar } |\tilde{\psi } (t) \rangle $. The time evolution equation for $|\tilde{\psi } (t) \rangle $ is
\begin{eqnarray}
i \hbar \frac{\partial }{\partial t} |\tilde{\psi } (t) \rangle  &=&  \tilde{V} (t) |\tilde{\psi } (t) \rangle ,
\end{eqnarray}
with
\begin{eqnarray}
\tilde{V} (t ) = e^{iH_0 t /\hbar} V e^{-i H_0 t/\hbar}  = g\sum_{n=-\infty}^\infty \left( |b\rangle \langle n |  e^{i (E_b - E_n)t /\hbar}+ {h.c.} \right) .
\end{eqnarray}
The time evolution operator in the interaction picture can be constructed formally as
\begin{eqnarray}\label{dyson}
\fl \quad \quad S(t) = I + \int_0^t d t_1 \left(\frac{1}{i\hbar} \tilde{V}(t_1 ) \right) +  \int_0^t d t_1 \int_{t_1}^t d t_2  \left(\frac{1}{i\hbar} \tilde{V}(t_2 ) \right) \left(\frac{1}{i\hbar} \tilde{V}(t_1 ) \right) + \cdots.\quad \;
\end{eqnarray}
This is the so-called Dyson series, in which the $s$th term corresponds to the $s$th term in (\ref{expan1}). The matrix element $S_{bb}(t)  \equiv \langle b | S(t)| b\rangle $ is of the form
\begin{eqnarray}\label{expan}
S_{bb}(t)  =\sum_{n =0}^\infty \left( \frac{g}{i\hbar}  \right )^{2n} C_n (t ),
\end{eqnarray}
with
\begin{eqnarray}\label{an}
 \fl \quad \quad \quad \quad C_n (t )&=& \sum_{m_1=-\infty}^\infty \cdots \sum_{m_n=-\infty}^\infty \int_0^{t} dt_1   \cdots \int_{t_{2n-1}}^t d t_{2n} e^{i \sum_{j=1}^n(E_b - E_{m_j})(t_{2j} -t_{2j-1})/\hbar} \nonumber\\
&=  &  \int_0^{t} dt_1 \cdots \int_{t_{2n-1}}^t d t_{2n}  \sum_{m_1=-\infty}^\infty \cdots \sum_{m_n=-\infty}^\infty e^{i \sum_{j=1}^n(E_b - E_{m_j})(t_{2j} -t_{2j-1})/\hbar }.\quad
\end{eqnarray}
Note that in (\ref{expan}) the odd terms in $g$ drop out. The reason is simply that the system has to jump even times between the discrete level and the quasi-continuum band to return to the discrete level.
Once we have calculated $ S_{bb}(t) $, we obtain the survival probability $P_i $ as
\begin{eqnarray}
 \fl \quad \quad  P_i = | \langle b |e^{-i H t /\hbar }|b \rangle |^2 = |\langle b | e^{- i H_0 t /\hbar } S(t) |b \rangle |^2 = |\langle b | e^{- i E_b t /\hbar } S(t) |b \rangle |^2 =|  S_{bb}(t) |^2.
\end{eqnarray}
It is ready to see that the dependence of $P_i $ on $E_b$ should be through the parameter $\alpha $.

In the second line of (\ref{an}), we need to do the summation
\begin{eqnarray}
\sum_{m=-\infty}^\infty e^{i m \Delta  t  /\hbar }.
\end{eqnarray}
Again, the Poisson summation formula is useful. The function relevant is $f(x) = e^{i   x }$, which is sampled with a period of $\Delta t /\hbar $. The Fourier transform of $f$ is
\begin{eqnarray}
F(q) = \int_{-\infty}^\infty d x e^{i  x } e^{- iq x } = 2 \pi \delta (q -1  ).
\end{eqnarray}
Therefore by the formula (\ref{momsum}) ($t_{H} \equiv 2\pi \hbar/\Delta  $),
\begin{eqnarray}\label{sum}
\sum_{m=-\infty}^\infty e^{i m \Delta t /\hbar   } = \frac{\hbar}{\Delta t } \sum_{m=-\infty}^\infty  2 \pi \delta \left(\frac{2\pi m \hbar}{\Delta t} - 1 \right ) = t_H \sum_{m=-\infty}^\infty \delta ( t- m t_{H} ).
\end{eqnarray}
Substituting (\ref{sum}) into (\ref{an}), we get
\begin{eqnarray}\label{an2}
 \fl C_n (t )  &= &    t_{H}^n \int_0^{t} dt_1  \cdots \int_{t_{2n-1}}^t d t_{2n}   \prod_{j=1}^n \left(  e^{i E_b (t_{2j} -t_{2j-1}) /\hbar}   \sum_{m_j=-\infty}^\infty  \delta [ (t_{2j} - t_{2j-1}) - m_j t_{H} ] \right )  \nonumber \\
\fl &= &   t_{H}^n  \int_0^{t} dt_1   \cdots \int_{t_{2n-1}}^t d t_{2n}   \prod_{j=1}^n \left(  e^{i E_b (t_{2j} -t_{2j-1})/\hbar }   \sum_{m_j=0}^\infty  \delta [(t_{2j} - t_{2j-1}) - m_j t_{H} ] \right ) \nonumber \\
\fl &=  &  t_{H}^n \sum_{m_1=0}^\infty  \cdots \sum_{m_n=0}^\infty  \int_0^{t} dt_1 \cdots \int_{t_{2n-1}}^t d t_{2n} \nonumber \\
 \fl & & \times \prod_{j=1}^n \left(  e^{i E_b (t_{2j} -t_{2j-1})/\hbar }  \delta [(t_{2j} - t_{2j-1}) - m_j t_{H} ] \right ) .
\end{eqnarray}
Now suppose $0< t< t_{{H}}$. In this case, $0\leq t_{2j} -t_{2j-1} < t_{H}$, and hence the only contributing term would be $m_j = 0$, $j = 1, \ldots, n $. Integrating out $t_{2j}$ and noting that $\int_0^\infty dt \delta (t) = 1/2 $, we get
\begin{eqnarray}\label{an3}
C_n (t )= t_{H}^n  \int_0^t d t_1 \int_{t_1}^t d t_3 \cdots \int_{t_{2n-3}}^t d t_{2n-1} \left( \frac{1}{2}\right)^n = \frac{1}{n!}\left( \frac{t_{H} t }{2 } \right)^n.
\end{eqnarray}
By (\ref{expan}), this leads to
\begin{eqnarray}\label{first}
S_{bb}(t)   = \exp \left(- \frac{\pi g^2  }{\Delta  \hbar} t\right ) = \exp\left(- \frac{\gamma  }{2} t \right),
\end{eqnarray}
where $\gamma \equiv  2 \pi g^2/\Delta \hbar $. It is purely exponential. Note that as in (\ref{linear}), there is no dependence on $\alpha $.

Next suppose $ t_{{H}} < t < 2 t_{{H}}$. In this case, besides the possibility of $m_j \equiv  0 $ for all $j$, one $m_j $ can be 1. The contribution of the first possibility to $C_n$ is given by (\ref{first}). In the second possibility, we make the change of variables, $(t_1, t_2, \ldots, t_{2n}) \rightarrow (s_1, s_2, \ldots, s_{2n})$,
\begin{eqnarray}
\cases{
s_i = t_i , &    if $ i \leq 2j -1 ,$ \\
s_i = t_i - t_{{H}}, & if $i > 2j - 1.$  \\ }
\end{eqnarray}
The Jacobian is apparently 1. The corresponding contribution to $C_n$, after integrating out $s_{2j}$, is ($\theta \equiv  2\pi \alpha $)
\begin{eqnarray}
 n t_{H}^n  \int_0^{t-t_{{H}}} d s_1  \cdots \int_{s_{2n-3}}^{t-t_{{H}} } d s_{2n-1} \left( \frac{1}{2}\right)^{n-1} e^{i E_b t_{{H}}/\hbar } =  \frac{ t_{H}^n (t-t_{{H}} )^n  }{(n-1)! 2^{n-1}} e^{i \theta }.
\end{eqnarray}
By (\ref{expan}), the corresponding contribution to $\langle b | S(t)| b\rangle $ is then
\begin{eqnarray}
-  \gamma  (t-t_{H}) e^{-\gamma (t-t_{H}) /2 } e^{i \theta  }.
\end{eqnarray}
In total, in the interval of $ t_{H} < t < 2 t_{H} $,
\begin{eqnarray}\label{second}
S_{bb}(t)   = e^{-\gamma t/2 } -   \gamma (t-t_{H} ) e^{-\gamma (t-t_{H} )/2 }  e^{i \theta  }.
\end{eqnarray}
Therefore, beyond $t_{H}$, $P_i $ is no longer purely exponential and is dependent on $\alpha$. Comparing (\ref{first}) and (\ref{second}), we see that $P_i$ has a cusp at $t = t_{H}$.

Next consider $2t_{H} < t< 3t_{H}  $. In this case, there are four types of contributing terms: (i) $m_j = 0$ for all $j$; (ii) $m_j=1$ for one $j$, while 0 for all other $j $; (iii) $m_j = 2$ for one $j$ while $0$ for all other $j$; (iv) $m_j = 1$ for two $j$, while $0$ for all other $j$. The contributions of (i) and (ii) to $ \langle b | S(t)| b\rangle $ are given by (\ref{first}) and (\ref{second}), respectively. The contribution of (iii) is given by
\begin{eqnarray}
\fl \quad \quad  \sum_{n=0}^\infty \left(-\frac{g^2}{\hbar^2} \right)^n t_{H}^n \left( \frac{1}{2}\right)^{n-1} e^{i 2 \theta  } \frac{(t-2t_{H} )^n}{n!} n = - \gamma (t-2t_{H} )  e^{-\gamma (t- 2t_{H} )/2} e^{i  2 \theta  }  .
\end{eqnarray}
The contribution of (iv) is
\begin{eqnarray}
\fl \quad \sum_{n=0}^\infty \left(-\frac{g^2}{\hbar^2} \right)^n t_{H}^n  \left( \frac{1}{2}\right)^{n-2} e^{i 2 \theta } \frac{(t-2t_{H})^n}{n!} \frac{n(n-1)}{2} = \frac{1}{2} \gamma^2  (t-2t_{{H}})^2 e^{-\gamma (t- 2t_{H}) /2} e^{i 2 \theta  }  .
\end{eqnarray}
In total, for $2t_{H} < t< 3t_{H}  $,
\begin{eqnarray}\label{third}
S_{bb}(t) &= &e^{-\gamma t/2 } -   \gamma (t-t_{H} ) e^{-\gamma (t-t_{H} ) /2} e^{i \theta  }
  \nonumber \\
  & & +\left[ \frac{1}{2} \gamma^2  (t-2t_{H} )^2 - \gamma (t-2t_{H} ) \right ] e^{-\gamma (t- 2t_{H} )/2} e^{i  2 \theta  }  .\;
\end{eqnarray}
Again, $P_i$ shows a cusp at $t = 2t_{H}$.

\begin{figure}[tb]
\includegraphics[width= 0.4\textwidth ]{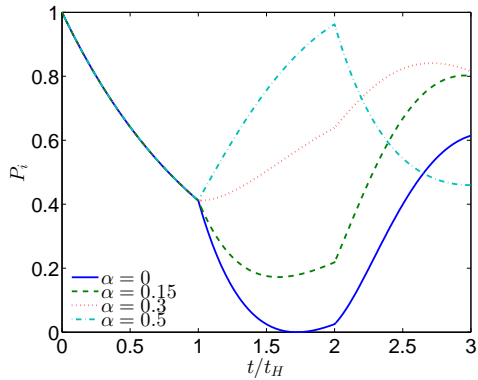}
\caption{(Color online) The survival probability $P_i $ on the discrete level $|b\rangle $ in the exact solution. It shows cusps periodically and the period is the Heisenberg time $t_{H} = 2\pi \hbar /\Delta $. In the first period, $P_i (t) = \exp(-\gamma t)$, with $\gamma = 2\pi  g^2 /\Delta \hbar  $. In this period, it is independent of the value of $\alpha $. But afterwards, it is sensitive to the value of $\alpha $. The parameters are chosen such that $g/ \Delta = 0.15 $.}
\label{exact}
\end{figure}

Similarly, we can determine the expression of $ S_{bb}(t)  $ for later intervals. For those interested, closed explicit expressions of $S_{bb}(t)$ for an arbitrary interval can be found in \cite{stey} and \cite{lefebvre}. But it is not hard to persuade oneself that in each interval, there will be extra terms contributing to $C_n$ than in the proceeding interval, hence extra terms in $S_{bb}(t)   $. Therefore, $P_i$ shows a cusp each time $t$ is an integral multiplier of the Heisenberg time $t_{H}$. In figure \ref{exact}, the trajectories of $P_i$ are shown for the same four values of $\alpha$ as in figure \ref{Wafig}. We see that when all orders of terms in (\ref{expan1}) are taken into account, many features in figure \ref{Wafig}, which contains only the first order term, are still preserved. For example, the periodic singularities persist and the period is unchanged, moreover, there is still no $\alpha $-dependence in the first interval $0 < t\leq t_{H}$.

The exact solution of the dynamics of the ideal model also sheds light on the theory of open systems. To model a system with a non-hermitian Hamiltonian or a Markov master equation, the common wisdom is that it should couple to sufficiently many levels. Now we see that another important factor is the level spacings. In the ideal model, in the first interval,  the discrete level is described exactly by the non-hermitian Hamiltonian $(E_b - i \gamma/2 )|b\rangle \langle b | $. However, the  incoherence and irreversibility is just apparent. After the Heisenberg time, as shown in figure \ref{exact}, the probability flow may turn around and significant revival could happen, which is a hallmark of coherence of the full system. To postpone the revival and make it irrelevant, the level spacing $\Delta $ has to go to zero (and the coupling $g$ should be diminished so as to keep $\gamma$ fixed). We note that this is implicitly assumed in the Weisskopf-Wigner theory of spontaneous decay \cite{scully}.

\section{Conclusions and discussions}

We consider the model [(\ref{modelh0}) and (\ref{modelv})] as of great pedagogical value. First, the model allows a lucid derivation of Fermi's golden rule. The calculation can be carried out straightforwardly without invoking any approximation or argument.
It illustrates the two key features of Fermi's golden rule very well, namely, the transition rate is proportional to the density of states and the coupling squared. Second, it demonstrates in a vivid way how the rule might break down beyond the Heisenberg time, a point less emphasized in quantum mechanics textbooks. One reason might be that many authors have in mind a true continuum, for which the Heisenberg time is infinite and thus irrelevant. However, a quasi-continuum with finite level spacings or finite Heisenberg times, is  a reality in many systems. This is the case, for example, in a molecule \cite{kyrola}, in a multi-mode optical cavity \cite{parker, meystre}, or in a one-dimensional tight binding model \cite{scienceopen}, where the piecewise linear behavior of the transition probability was observed.

Interestingly, the transition dynamics of the model can actually be solved exactly by collecting all orders of terms in the Dyson series, which in many quantum mechanics textbooks is introduced formally but not put into full use.
The rigorous exponential decay in the first interval is rare instead of common. The periodic cusps imply that the kinks in the first order perturbation approximation is not mere artifact; they are modified but not smeared out completely in the exact solution. Recently, cusps (periodic or not) have been found also in the quench dynamics of the transverse Ising model \cite{heyl}, the tight binding model \cite{bloch, bloch2}, and some non-integrable models \cite{schuricht}. The mechanism varies from case to case. In view of these new progresses, why and when can the time evolution of some quantity be singular is now a problem worth consideration.

In one word, we think the model deserves to be introduced into quantum mechanics textbooks or lectures. Although the derivation of Fermi's golden rule for this special model is no substitute of the more general and more conventional one, it can at least be incorporated into the homework session. It surely will give the students a feeling of the contents and limitations of the rule. In addition, its exact solution can also show the students that the Dyson series, which seems just a formal object in many textbooks, can actually be summed up  in a very neat form in some particular cases.

Finally, to be fair, we have to mention that decay of a discrete level into a continuum is a fundamental problem in quantum mechanics
and a more systematic treatment is by the resonant state theory \cite{hunziker}, which requires mathematical knowledge far beyond that is displayed here. The point is that, the ideal model can have variants, many of which fall beyond the golden rule paradigm. For example, what if the coupling strength $g_n$ is non-uniform and vanishes at $E_b$ \cite{g0}? The formula (\ref{fgr}) predicts a zero transition rate, but the transition is still to occur. This just means that in this case, Fermi's golden rule is inadequate for predicting the decay behavior of the discrete level. As yet another scenario, the continuum band might have a lower threshold  (as in any realistic system) and the discrete level can be at or near the threshold \cite{dinu, jensen}. In this case, the fortunate simplification we have here by the Poisson summation formula is missing. Again, more sophisticated analysis shows that the linear behavior (\ref{fgr}) and the exponential behavior (\ref{first}) may be untrue or have to be modified in a nontrivial way \cite{dinu, jensen}.

\section*{Acknowledgments}

The author is grateful to Hua-Tong Yang, Daniel Braak and Quan-Hui Liu for their helpful comments. This work is supported by the Fujian Provincial Science Foundation under grant number 2016J05004, the China Postdoctoral Science Foundation, National High Technology Research and Development Program of China under Grant 2015AA01A304, and the Foundation of LCP.


\begin{thebibliography}{59}
\bibitem{around}
It is actually the other way around, i.e., the Fourier transform of the window function is the $\sinc $ function.
\bibitem{exponentialtype}
A function $f(z)$ defined on the complex plane is said to be of exponential type if there exist constants $ A$  and $\alpha $ such that $|f(z )| \leq A e^{\alpha  |z|}$ in the limit of $|z| \rightarrow \infty $. Letting $\beta  $ stand for the infimum of all such $\alpha $, one then says that the function $ f $ is of exponential type $\beta  $.
\end{thebibliography}
\end{document}